\documentclass{article}
\usepackage{amsfonts}

\input{tcilatex}

\begin{document}

{\footnotesize {Mathematical Problems of Computer Science {\small 26, 2006,
28--32.}}}

\bigskip

\bigskip

\bigskip

\begin{center}
{\Large \textbf{On Interval Colorings of Complete }}$k-${\Large \textbf{%
partite Graphs }}$K_{n}^{k}$

{\normalsize Rafael R. Kamalian and Petros A. Petrosyan}

{\small Institute for Informatics and Automation Problems (IIAP) of NAS of RA%
}

{\small e-mails rrkamalian@yahoo.com, pet\_petros@yahoo.com}

\bigskip

\textbf{Abstract}
\end{center}

Problems of existence, construction and estimation of parameters of interval
colorings of complete $k-$partite graphs $K_{n}^{k}$ are investigated.

\bigskip

Let $G=(V,E)$ be an undirected graph without loops and multiple edges [1], $%
V(G)$ and $E(G)$ be the sets of vertices and edges of $G$, respectively. The
degree of a vertex $x\in V(G)$ is denoted by $d_{G}(x)$, the maximum degree
of a vertex of $G$-by $\Delta (G)$, and the chromatic index [2] of $G$-by $%
\chi ^{\prime }(G)$. A graph is regular, if all its vertices have the same
degree. If $\alpha $ is a proper edge coloring of the graph $G$ [3], then
the color of an edge $e\in E(G)$ in the coloring $\alpha $ is denoted by $%
\alpha (e)$, and the set of colors of the edges that are incident to a
vertex $x\in V(G)$ , is denoted by $S(x,\alpha )$. For a non-empty subset $D$
of $%
%TCIMACRO{\U{2124} }%
%BeginExpansion
\mathbb{Z}
%EndExpansion
_{+}$, let $l\left( D\right) $ and $L\left( D\right) $ be the minimal and
maximal element of $D$, respectively. A non-empty subset $D$ of $%
%TCIMACRO{\U{2124} }%
%BeginExpansion
\mathbb{Z}
%EndExpansion
_{+}$ is interval, if $l\left( D\right) \leq t\leq L\left( D\right) ,$ $t\in 
$ $%
%TCIMACRO{\U{2124} }%
%BeginExpansion
\mathbb{Z}
%EndExpansion
_{+}$ implies that $t\in D$. Interval $D$ is referred to be $\left(
q,h\right) $-interval if $l\left( D\right) =q,$ $\left\vert D\right\vert =h$
and is denoted by $Int\left( q,h\right) $. For intervals $D_{1}$ and $D_{2}$
with $\left\vert D_{1}\right\vert =\left\vert D_{2}\right\vert =h$ , and a $%
p\in 
%TCIMACRO{\U{2124} }%
%BeginExpansion
\mathbb{Z}
%EndExpansion
_{+}$, the notation $D_{1}\oplus p=$ $D_{2}$ means: $l\left( D_{1}\right)
+p=l\left( D_{2}\right) $.

A proper coloring $\alpha $ of edges of $G$ with colors $1,2,\ldots ,t$ is
an interval $t$-coloring of $G$ [4], if for each color $i,1\leq i\leq t,$
there exists at least one edge $e_{i}\in E(G)$ with $\alpha (e_{i})=i$, and
the edges incident with each vertex $x\in V(G)$ are colored by $d_{G}(x)$
consecutive colors.

A graph $G$ is interval colorable, if there is $t\geq 1$, for which $G$ has
an interval $t$-coloring.

The set of all interval colorable graphs is denoted by $\mathcal{N}$ \ [5].

For $G\in \mathcal{N}$ the least and the greatest values of$\ t$, for which $%
G$ has an interval $t$-coloring, is denoted by $w(G)$ and $W(G)$,
respectively.

In [5] it is proved:

\textbf{Theorem 1}. Let $G$ be a regular graph.

1) $G\in \mathcal{N}$ iff $\chi ^{\prime }(G)=\Delta (G)$.

2) If $G\in \mathcal{N}$ and $\Delta (G)\leq t\leq W(G)$, then $G$ has an
interval $t$-coloring.

\textbf{Theorem 2 }[6]. Let $n=p\cdot 2^{q}$, where $p$ is odd, and $q\in 
%TCIMACRO{\U{2124} }%
%BeginExpansion
\mathbb{Z}
%EndExpansion
_{+}$. Then $W\left( K_{2n}\right) \geq 4n-2-p-q$.

In this paper interval colorings of complete $k$-partite graphs $K_{n}^{k}$
[7] are investigated, where$\ $

\begin{center}
$V\left( K_{n}^{k}\right) =\left\{ x_{j}^{(i)}|\text{ }1\leq i\leq k,1\leq
j\leq n\right\} $, $E\left( K_{n}^{k}\right) =\left\{ \left(
x_{p}^{(i)},x_{q}^{(j)}\right) |\text{ }1\leq i<j\leq k,1\leq p\leq n,1\leq
q\leq n\right\} $.
\end{center}

It is not hard to see that $\Delta (K_{n}^{k})=\left( k-1\right) \cdot n$.

From the results of [8] we imply that

\begin{center}
$\chi ^{\prime }(K_{n}^{k})=\left\{ 
\begin{array}{lll}
\left( k-1\right) \cdot n, & \text{if} & n\cdot k\text{ is even,} \\ 
\left( k-1\right) \cdot n+1, & \text{if} & n\cdot k\text{ is odd.}%
\end{array}%
\right. $
\end{center}

Theorem 1 implies:

\textbf{Corollary\ 1.}

1) $K_{n}^{k}\in \mathcal{N}$, if $n\cdot k$ is even;

2) $K_{n}^{k}\notin \mathcal{N}$, if $n\cdot k$ is odd.

\textbf{Corollary\ 2.} If $n\cdot k$ is even, then $w(K_{n}^{k})=\left(
k-1\right) \cdot n$.

\textbf{Theorem 3. }If $k$ is even, then\ $W\left( K_{n}^{k}\right) \geq
\left( \frac{3}{2}k-1\right) \cdot n-1$.

\textbf{Proof.} Let $V\left( K_{n}^{k}\right) =\left\{ x_{j}^{(i)}|\text{ }%
1\leq i\leq k,1\leq j\leq n\right\} $, \newline
$E\left( K_{n}^{k}\right) =\left\{ \left( x_{p}^{(i)},x_{q}^{(j)}\right) |%
\text{ }1\leq i<j\leq k,1\leq p\leq n,1\leq q\leq n\right\} $.

For the graph $K_{n}^{k}$ define an edge coloring $\lambda $ as follows:

\bigskip for $\ i=1,...,\left\lfloor \frac{k}{{\large 4}}\right\rfloor ,$ $%
j=2,...,\frac{k}{2},$ $i<j,$\ $i+j\leq \frac{k}{2}+1,$ $p=1,...,n,$ $%
q=1,...,n$ set:

\begin{center}
$\lambda \left( \left( x_{p}^{(i)},x_{q}^{(j)}\right) \right) =\left(
i+j-3\right) \cdot n+p+q-1;$\ \ \ \ 

\ \ 
\end{center}

for \ $i=2,...,\frac{k}{2}-1,$ $j=\left\lfloor \frac{k}{{\large 4}}%
\right\rfloor +2,...,\frac{k}{2},$ $i<j,$\ $i+j\geq \frac{k}{2}+2,$ $%
p=1,...,n,$ $q=1,...,n$ set:$\bigskip $

\begin{center}
$\lambda \left( \left( x_{p}^{(i)},x_{q}^{(j)}\right) \right) =\left( i+j+%
\frac{k}{2}-4\right) \cdot n+p+q-1;$\ \ \ 

\ \ 
\end{center}

\ \bigskip for \ $i=3,...,\frac{k}{2},$ $j=\frac{k}{2}+1,...,k-2,$ $j-i\leq 
\frac{k}{2}-2,$ $p=1,...,n,$ $q=1,...,n$ set:

\begin{center}
$\lambda \left( \left( x_{p}^{(i)},x_{q}^{(j)}\right) \right) =\left( \frac{k%
}{2}+j-i-1\right) \cdot n+p+q-1;$\ \ \ 

\ \ 
\end{center}

for \ $i=1,...,\frac{k}{2},$ $j=\frac{k}{2}+1,...,k,$ $j-i\geq \frac{k}{2},$ 
$p=1,...,n,$ $q=1,...,n$ set:\ \ \ 

\begin{center}
\ \ 

$\lambda \left( \left( x_{p}^{(i)},x_{q}^{(j)}\right) \right) =\left(
j-i-1\right) \cdot n+p+q-1;$\ \ \ 

\ \ 
\end{center}

for \ $i=2,...,1+\left\lfloor \frac{k{\large -2}}{{\large 4}}\right\rfloor ,$
$j=\frac{k}{2}+1,...,\frac{k}{2}$ $+\left\lfloor \frac{k{\large -2}}{{\large %
4}}\right\rfloor ,$ $\ j-i=\frac{k}{2}-1,$ $p=1,...,n,$ $q=1,...,n$ \
set:\bigskip

\begin{center}
\bigskip $\lambda \left( \left( x_{p}^{(i)},x_{q}^{(j)}\right) \right)
=\left( 2i-3\right) \cdot n+p+q-1;$\ \ 
\end{center}

for \ $i=$ $\left\lfloor \frac{k{\large -2}}{{\large 4}}\right\rfloor +2,...,%
\frac{k}{2},$ $\ j=\frac{k}{2}+1+\left\lfloor \frac{k{\large -2}}{{\large 4}}%
\right\rfloor ,...,k-1,$ $\ j-i=\frac{k}{2}-1,$ $p=1,...,n,$ $q=1,...,n$ \
set:\bigskip

\begin{center}
\ $\lambda \left( \left( x_{p}^{(i)},x_{q}^{(j)}\right) \right) =\left(
i+j-3\right) \cdot n+p+q-1;$\ \ \ \ 

\ \ 
\end{center}

for \ $i=\frac{k}{2}+1,...,\frac{k}{2}+\left\lfloor \frac{k}{{\large 4}}%
\right\rfloor -1,$ $j=\frac{k}{2}+2,...,k-2,$ $i<j,$\ $i+j\leq \frac{3}{2}%
k-1,$ $p=1,...,n,$ $q=1,...,n$ \ set:

\begin{center}
$\lambda \left( \left( x_{p}^{(i)},x_{q}^{(j)}\right) \right) =\left(
i+j-k-1\right) \cdot n+p+q-1;$
\end{center}

for \ $i=\frac{k}{2}+1,...,k-1,$ $j=\frac{k}{2}+\left\lfloor \frac{k}{4}%
\right\rfloor +1,...,k,$ $i<j,$\ $i+j\geq \frac{3}{2}k,$ $p=1,...,n,$ $%
q=1,...,n$ set:\bigskip

\begin{center}
\bigskip $\lambda \left( \left( x_{p}^{(i)},x_{q}^{(j)}\right) \right)
=\left( i+j-\frac{k}{2}-2\right) \cdot n+p+q-1.$
\end{center}

Let us show that $\lambda $ is an interval $\left( \left( \frac{3}{2}%
k-1\right) \cdot n-1\right) -$coloring of the graph $K_{n}^{k}$.

First of all let us show that for $i=1,2,...,\left( \frac{3}{2}k-1\right)
\cdot n-1$ there is an edge $e_{i}$\ $\in E\left( K_{n}^{k}\right) $ such
that $\lambda \left( e_{i}\right) =i$.

Consider the vertices $%
x_{1}^{(1)},x_{2}^{(1)},...,x_{n}^{(1)},x_{1}^{(k)},x_{2}^{(k)},...,x_{n}^{(k)} 
$. It is not hard to see that for $j=1,2,...,n$\bigskip

\begin{center}
$S\left( x_{j}^{(1)},\lambda \right) =\dbigcup\limits_{l=1}^{k-1}\left(
Int\left( j,n\right) \oplus n\cdot \left( l-1\right) \right) $ \ and \ $%
S\left( x_{j}^{(k)},\lambda \right) =\dbigcup\limits_{l=\frac{k}{2}}^{\frac{3%
}{2}k-2}\left( Int\left( j,n\right) \oplus n\cdot \left( l-1\right) \right)
. $\bigskip
\end{center}

Let $\overline{C}$ and $\overline{\overline{C}}-$be the subsets of colors of
the edges, that are incident to the vertices $%
x_{1}^{(1)},x_{2}^{(1)},...,x_{n}^{(1)}$ and $%
x_{1}^{(k)},x_{2}^{(k)},...,x_{n}^{(k)}$ in a coloring $\lambda $,
respectively, that is:

\begin{center}
\bigskip $\overline{C}=\dbigcup\limits^{n}_{j=1}S\left( x_{j}^{(1)},\lambda
\right) $ and \ $\overline{\overline{C}}=\dbigcup\limits^{n}_{j=1}S\left(
x_{j}^{(k)},\lambda \right) $.
\end{center}

It is not hard to see that $\overline{C}\cup \overline{\overline{C}}=\left\{
1,2,\ldots ,\left( \frac{3}{2}k-1\right) \cdot n-1\right\} $, and,
therefore, for $i=1,2,...,\left( \frac{3}{2}k-1\right) \cdot n-1$ there is
an edge $e_{i}\in E\left( K_{n}^{k}\right) $ such that $\lambda \left(
e_{i}\right) =i$.

Now, let us show that the edges that are incident to a vertex $\ v\in
V\left( K_{n}^{k}\right) $ are colored by $\left( k-1\right) \cdot n$
consecutive colors.

Let $x_{j}^{(i)}\in $ $V\left( K_{n}^{k}\right) ,$ where $1\leq i\leq
k,1\leq j\leq n$.

Case 1. $1\leq i\leq 2,1\leq j\leq n$.

It is not hard to see that

\begin{center}
\bigskip $S\left( x_{j}^{(1)},\lambda \right) =S\left( x_{j}^{(2)},\lambda
\right) =\dbigcup\limits_{l=1}^{k-1}\left( Int\left( j,n\right) \oplus
n\cdot \left( l-1\right) \right) =Int\left( j,\left( k-1\right) \cdot
n\right) $.
\end{center}

Case 2. $3\leq i\leq \frac{k}{2},1\leq j\leq n$.

It is not hard to see that

\begin{center}
\bigskip $S\left( x_{j}^{(i)},\lambda \right)
=\dbigcup\limits_{l=i-1}^{k-3+i}\left( Int\left( j,n\right) \oplus n\cdot
\left( l-1\right) \right) =Int\left( j+n\cdot \left( i-2\right) ,\left(
k-1\right) \cdot n\right) $.\bigskip
\end{center}

Case 3.\ $\frac{k}{2}+1\leq i\leq k-2,1\leq j\leq n$.

It is not hard to see that

\begin{center}
\bigskip $S\left( x_{j}^{(i)},\lambda \right) =\dbigcup\limits_{l=i-\frac{k}{%
2}+1}^{\frac{k}{2}-1+i}\left( Int\left( j,n\right) \oplus n\cdot \left(
l-1\right) \right) =Int\left( j+n\cdot \left( i-\frac{k}{2}\right) ,\left(
k-1\right) \cdot n\right) $.
\end{center}

Case 4. $k-1\leq i\leq k,1\leq j\leq n$.

It is not hard to see that

\begin{center}
\bigskip $S\left( x_{j}^{(k-1)},\lambda \right) =S\left( x_{j}^{(k)},\lambda
\right) =\dbigcup\limits^{\frac{3}{2}k-2}_{l=\frac{k}{2}}\left( Int\left(
j,n\right) \oplus n\cdot \left( l-1\right) \right) =Int\left( j+n\cdot
\left( \frac{k}{2}-1\right) ,\left( k-1\right) \cdot n\right) $.
\end{center}

\textbf{Theorem 3} is proved.

\textbf{Corollary\ 3. }If $k$ is even and $\left( k-1\right) \cdot n\leq
t\leq $\ $\left( \frac{3}{2}k-1\right) \cdot n-1$, then $K_{n}^{k}$ has an
interval \ $t-$coloring.

\textbf{Theorem 4. }Let $k=p\cdot 2^{q}$, where\ $p$ is odd, and $q\in 
%TCIMACRO{\U{2115} }%
%BeginExpansion
\mathbb{N}
%EndExpansion
$. Then \ $W\left( K_{n}^{k}\right) \geq \left( 2k-p-q\right) \cdot n-1$.

\textbf{Proof.} Let $V\left( K_{n}^{k}\right) =\left\{ x_{j}^{(i)}|\text{ }%
1\leq i\leq k,1\leq j\leq n\right\} $,

$E\left( K_{n}^{k}\right) =\left\{ \left( x_{r}^{(i)},x_{s}^{(j)}\right) |%
\text{ }1\leq i<j\leq k,1\leq r\leq n,1\leq s\leq n\right\} $.

Consider the graph $K_{k}$, where\ $V\left( K_{k}\right) =\left\{
u_{1},u_{2},\ldots ,u_{k}\right\} $,

$E\left( K_{k}\right) =\left\{ \left( u_{i},u_{j}\right) \text{ }|\text{ }%
1\leq i<j\leq k\right\} $. Theorem 2 implies that if $k=p\cdot 2^{q}$,
where\ $p\ $is odd, and $q\in 
%TCIMACRO{\U{2115} }%
%BeginExpansion
\mathbb{N}
%EndExpansion
$, then $W\left( K_{k}\right) \geq 2k-1-p-q.$ Suppose $\varphi $ is an
interval $\left( 2k-1-p-q\right) -$coloring of $K_{k}$.

Define a coloring $\psi $ of the edges of $K_{n}^{k}$ as follows:

For $\ i=1,2,\ldots ,k$ and $\ j=1,2,\ldots ,k$ , $\ i\neq j$ set:\bigskip

\begin{center}
$\psi \left( \left( x_{r}^{(i)},x_{s}^{(j)}\right) \right) =\left( \varphi
\left( \left( u_{i},u_{j}\right) \right) -1\right) \cdot n+r+s-1$,
\end{center}

where\ \ $r=1,2,...,n,$ $s=1,2,...,n$.\bigskip

Let us show that $\psi $ is an interval $\left( \left( 2k-p-q\right) \cdot
n-1\right) -$coloring of the graph $K_{n}^{k}$.

The definition of \ $\psi $ and the equalities \ $L\left( S\left(
u_{i},\varphi \right) \right) -$\ $l\left( S\left( u_{i},\varphi \right)
\right) =k-2$, $\ i=1,2,\ldots ,k$ imply that:

1) \ $S\left( x_{j}^{(i)},\psi \right) =\dbigcup\limits_{m=l\left( S\left(
u_{i},\varphi \right) \right) }^{L\left( S\left( u_{i},\varphi \right)
\right) }\left( Int\left( j,n\right) \oplus n\cdot \left( m-1\right) \right)
=$ \newline

$\bigskip $

$=Int\left( j+n\cdot \left( l\left( S\left( u_{i},\varphi \right) \right)
-1\right) ,\left( k-1\right) \cdot n\right) $

\bigskip

for $\ i=1,2,\ldots ,k$\ and $\ j=1,2,\ldots ,n$;\bigskip

2) \ $\dbigcup\limits_{i=1}^{k}\dbigcup\limits_{j=1}^{n}$\ $S\left(
x_{j}^{(i)},\psi \right) =Int\left( 1,\left( 2k-p-q\right) \cdot n-1\right) $%
.\bigskip

This show that $\psi $ is an interval $\left( \left( 2k-p-q\right) \cdot
n-1\right) -$coloring of the graph $K_{n}^{k}$.

\textbf{Theorem 4} is proved.

\textbf{Corollary\ 4. }Let $k=p\cdot 2^{q}$, where \ $p\ $ is odd and $q\in 
%TCIMACRO{\U{2115} }%
%BeginExpansion
\mathbb{N}
%EndExpansion
$. If $\left( k-1\right) \cdot n\leq t\leq $\ $\left( 2k-p-q\right) \cdot
n-1 $, then $K_{n}^{k}$ has an interval \ $t$-coloring.

\begin{center}
\bigskip
\end{center}

\end{document}